\documentclass[apj]{emulateapj}

\usepackage{url,amsfonts,epsfig}
\usepackage{graphicx}
\usepackage{amsmath}
\usepackage{amssymb}
\usepackage[section]{placeins}
\usepackage{chngpage}
\usepackage{calc}
\usepackage{subfigure}
\usepackage{mathtools}

\usepackage{natbib}
\makeatletter 
\renewcommand\@biblabel[1]{}

\shorttitle{Orbital evolution of mass-transferring eccentric binary systems}
\shortauthors{Dosopoulou and Kalogera}
\begin{document}
\def\gap{\;\rlap{\lower 2.5pt
\hbox{$\sim$}}\raise 1.5pt\hbox{$>$}\;}
\def\lap{\;\rlap{\lower 2.5pt
 \hbox{$\sim$}}\raise 1.5pt\hbox{$<$}\;}

\newcommand\sbh{MBH}
\newcommand\NSC{NSC}
\title{Orbital evolution of mass-transferring eccentric binary systems.\\ II. Secular Evolution}

\author{Fani Dosopoulou and Vicky Kalogera}
\email{FaniDosopoulou2012@u.northwestern.edu}
\affil{Center for Interdisciplinary Exploration and Research in Astrophysics (CIERA) and Department of Physics and Astrophysics,
Northwestern University, Evanston, IL 60208}

\begin{abstract}
Finite eccentricities in mass-transferring eccentric binary systems can be explained by taking into account mass-loss and mass-transfer processes that often occur in these systems. These processes can be treated as perturbations to the general two-body problem. The time-evolution equations for the semi-major axis and the eccentricity derived from perturbative methods are in general phase-dependent. The osculating semi-major axis and eccentricity change over the orbital timescale and they are not easy to implement in binary evolution codes like MESA. However, the secular orbital element evolution equations can be simplified averaging over the rapidly varying true anomalies. In this paper, we derive the secular time-evolution equations for the semi-major axis and the eccentricity for various mass-loss/transfer processes using either the adiabatic approximation or the assumption of delta-function mass-loss/transfer at periastron. We begin with the cases of isotropic and anisotropic wind mass-loss. We continue with conservative and non-conservative non-isotropic mass ejection/accretion (including RLOF) for both point-masses and extended bodies. We conclude with the case of phase-dependent mass accretion. Comparison of the derived equations with similar work in the literature is included and explanation of the existing discrepancies is provided.
 \end{abstract}
 

\section{Introduction}

The evolution of stellar binaries can be affected by a variety of physical phenomena, including mass-loss and mass-transfer processes. Such processes affect the orbital evolution of these systems and thus consist an important part of the binary evolution theory. In Dosopoulou $\&$ Kalogera 2016 (from now on Paper I), we treated different cases of mass-loss/transfer as perturbations to the general two-body problem. In that paper, we described the general perturbation method that can be used to calculate the orbital evolution of the binary and using this method we derived the general phase-dependent time-evolution equations for the semi-major axis and the eccentricity. Here, assuming that the binary is slowly evolving we orbit-average the phase-dependent time-evolution equations and derive the secular time-evolution equations for the semi-major axis and the eccentricity. The secular evolution equations are derived both in the adiabatic regime as well as under the assumption of delta-function perturbation.
\par We consider two bodies of masses $M_{1}$ and $M_{2}$ in an eccentric binary system of total mass $M$. We refer to these as the donor and the accretor, respectively. The donor $M_{1}$ is losing mass with a rate $\dot{M}_{1}$ and the accretor is accreting mass with a rate $\dot{M}_{2}$ so that the total mass-loss rate is given by $\dot{M}=\dot{M}_{1}+\dot{M}_{2}$. For non-conservative mass-transfer, only a fraction $\gamma$ of the lost mass is accreted. This means that in this case we have $\dot{M}_{2}=-\gamma \dot{M}_{1}$, where $\dot{M}_{1} < 0$. For the general case of non-conservative mass-transfer, we derived in Paper I the following phase-dependent time-evolution equations of the osculating semi-major axis $a$ and eccentricity $e$ in the $K_{R}$ reference system [see Figure $1$ here and equations $(117)$ and $(118)$ in Paper I]
{\setlength{\abovedisplayskip}{4pt}
\setlength{\belowdisplayskip}{4pt}
\begin{align}
 \nonumber\left( \frac{da}{dt}\right)_{non-con}&=\frac{2}{n\sqrt{1-e^{2}}}\left[b_{r}e\sin{\nu}
+b_{\tau}(1+e\cos{\nu})\right]\\ \nonumber
& -2ha\left[ \frac{e^{2}+2e\cos{\nu}+1}{1-e^{2}}\right]\\
&-(\zeta+1)\left[ \frac{e^{2}+2e\cos{\nu}+1}{1-e^{2}}\right]\frac{\dot{M}}{M}\label{agen}\\
\begin{split}
\left(\frac{de}{dt}\right)_{non-con}&=\frac{(1-e^{2})^{1/2}}{na}\left[b_{r}\sin{\nu}
\right. \\
&\left. +b_{\tau}\frac{2\cos{\nu}+e(1+\cos^{2}\nu)}{1+e\cos{\nu}}\right]\\
&-2h\left[e+\cos{\nu}\right]-(\zeta+1)\left[e+\cos{\nu}\right]\frac{\dot{M}}{M} \label{egen}
\end{split}
\end{align}}
where the subscripts $r$ and $\tau$ refer to the radial and tangential components respectively, $\nu$ is the true anomaly and $\zeta$ is the fraction of the total orbital angular momentum removed from the system. The other parameters are defined by
{\setlength{\abovedisplayskip}{8pt}
\setlength{\belowdisplayskip}{0pt}
\begin{align}
\begin{split}
\bold{b}&\equiv \dot{M}_{2}\left( \frac{\bold{w}_{2}}{M_{2}}
+\frac{\bold{\omega}_{orb}\times \bold{r}_{A_{2}}}{M_{2}}
+\frac{\bold{w}_{1}}{M_{1}}\right.\\ &\left. 
+\frac{\bold{\omega}_{orb}\times \bold{r}_{A_{1}}}{M_{1}}
\right)+\ddot{M}_{2}\left(\frac{\bold{r}_{A_{2}}}{M_{2}}
+\frac{\bold{r}_{A_{1}}}{M_{1}}\right)
\end{split}\label{generalb}
\end{align}
}
and
{\setlength{\abovedisplayskip}{6pt}
\setlength{\belowdisplayskip}{0pt}
\begin{align}
h&\equiv\dot{M}_{2}\left(\frac{1}{M_{2}}-\frac{1}{M_{1}}\right)\label{h2}
\end{align}}
{\setlength{\abovedisplayskip}{-8pt}
\setlength{\belowdisplayskip}{0pt}
\begin{align}
\dot{M}_{2}&=\dot{M}_{2,acc}=-\gamma \dot{M}_{1}\label{extrah}\\
\dot{M}_{1}&=-\dot{M}_{2,acc}=\dot{M}_{1,acc}=\gamma \dot{M}_{1}\\
\dot{M}&=(1-\gamma)\dot{M}_{1}\label{lastt}.
\end{align}}
\par In equations $(\ref{generalb})-(\ref{lastt})$, $\bold{w}_{1}$ and $\bold{w}_{2}$ are the absolute velocities of the ejected and accreted mass respectively and $\bold{r}_{A_{1}}$ and $\bold{r}_{A_{1}}$ are the positions of the ejection and accretion points relative to the bodies centers-of-mass. Furthermore, $\bold{v}_{1}$ and $\bold{v}_{2}$ are the orbital velocities of the two bodies with respect to the inertial frame and $\bold{\omega}_{orb}$ is the orbital angular frequency.
\par In the case the bodies in the binary system can be treated as point-masses we can simplify equations  $(\ref{agen})$ and $(\ref{egen})$ and rewrite them as [see equations $(123)$ and $(124)$ in Paper I]
{\setlength{\abovedisplayskip}{6pt}
\setlength{\belowdisplayskip}{8pt}
\begin{align}
 \nonumber \left(\frac{da}{dt}\right)_{non-con}^{point}&=\frac{2}{n\sqrt{1-e^{2}}}\left[c_{r}e\sin{\nu}
+c_{\tau}(1+e\cos{\nu})\right]\\
&-(\zeta+1)\left[ \frac{e^{2}+2e\cos{\nu}+1}{1-e^{2}}\right]\frac{\dot{M}}{M}\label{agenpoint}\\
\begin{split}
\left(\frac{de}{dt}\right)_{non-con}^{point}&=\frac{(1-e^{2})^{1/2}}{na}\left[c_{r}\sin{\nu}
\right. \\
&\left. +c_{\tau}\frac{2\cos{\nu}+e(1+\cos^{2}\nu)}{1+e\cos{\nu}}\right]\\
&-(\zeta+1)\left[e+\cos{\nu}\right]\frac{\dot{M}}{M}.\label{egenpoint}
\end{split}
\end{align}}
where we make use of the definition
 \begin{equation}
\bold{c}\equiv\frac{\dot{M_{2}}}{M_{2}}\left( \bold{w}_{2}-\bold{v}_{2}\right)
-\frac{\dot{M_{1}}}{M_{1}}\left( \bold{w}_{1}-\bold{v}_{1}
 \right)\label{point1}
\end{equation}
and the subscripts $r$ and $\tau$ refer to the radial and tangential components, respectively.
\par Equations  $(\ref{agen})$ and $(\ref{egen})$ are in general phase-dependent and indicate that the semi-major axis and the eccentricity change over orbital timescales undergoing periodic oscillations. This makes these equations difficult to implement in binary evolution codes like StarTrack \citep{2008ApJS..174..223B}, BSE \citep{2002MNRAS.329..897H} or MESA \citep{2011ApJS..192....3P,2013ApJS..208....4P,2015ApJS..220...15P}. This difficulty emerges from the fact that the semi-major axis and eccentricity change on an orbital timescale $\tau_{orb}$ while in the stellar evolution codes mentioned before the time-step is limited by the thermal timescale $\tau_{th}$, where $\tau_{th}>>\tau_{orb}$. This means that these codes cannot keep track of the evolutionary changes in the orbital elements using equations  $(\ref{agen})$ and $(\ref{egen})$. However, equations $(\ref{agen})$ and $(\ref{egen})$ can be simplified if they are orbit-averaged - a procedure that can smooth out the aforementioned oscillations. When the assumption of slowly evolving systems can be made, the orbit-averaged equations provide information about the secular evolution of the orbital elements and from now on we are going to refer to these equations as \emph{secular time-evolution equations}. The secular time-evolution equations can then be implemented in the stellar evolution codes mentioned before. In this paper, we perform the orbit-averaging process either under the adiabatic approximation or considering delta-function mass-loss/transfer at periastron. Under either of these assumptions, the derived secular time-evolution equations for the semi-major axis and the eccentricity are simplified further and can be used to efficiently model interacting eccentric binary systems.
\par This paper is organized as follows. In Section 2 we present the orbit-averaging process and describe as well as comment on the adiabatic approximation and the assumption of delta-function perturbation. In Section 3 we discuss briefly the importance of the reference frame choice and derive for two different reference frames the secular orbital element time-evolution equations both in the adiabatic regime and for delta-function perturbation under the additional assumption of a phase-independent perturbation. In Section 4 we derive the secular time-evolution equations for the semi-major axis and the eccentricity both in the adiabatic regime and under the assumption of delta-function mass-loss/transfer at periastron for the case of isotropic wind mass-loss. In Section 5 we apply the results of Section 3 in the case of anisotropic wind mass-loss and discuss qualitatively the effect of some simple wind structures on the binary orbital evolution, including jets. In Section 6 we study the non-isotropic ejection and accretion in a binary for the cases of both conservative and non-conservative mass-transfer (including Roche-Lobe-Overflow (RLOF)) assuming extended bodies. In Section 7 we derive the secular orbital element time-evolution equations for some commonly used characteristic examples of mass-loss/transfer under the additional point-mass approximation. Section 8 deals with the case of a phase-dependent mass accretion rate (Bondi-Hoyle scenario) and in Section 9 we summarize our results.

\section{Secular Evolution}

The time-evolving orbital elements of a perturbed two-body system undergo physical oscillations due to the periodicity of the orbit [see equations $(\ref{agen})$ and $(\ref{egen})$]. These oscillations can be smoothed out adopting orbit-averaging techniques that provide information about the secular evolution of the system. Following Lagrange planetary equations (which refer to osculating orbital elements) for a general perturbation [see equations $(18)-(23)$ in Paper I] the time-evolution equation for the true anomaly $\nu$ is given by 
\begin{align}
\frac{d\nu}{dt}&=\left(\frac{d\nu}{dt}\right)_{unperturbed}+\left(\frac{d\nu}{dt}\right)_{perturbation}\label{true}\\
&=\frac{n(1+e\cos{\nu})^{2}}{(1-e^{2})^{3/2}}-\frac{d\omega}{dt}-\cos{i}\frac{d\Omega}{dt}\label{true2}
\end{align}
where $\omega$ is the argument of periapsis and $\Omega$ is the longitude of the ascending node. For a general perturbation, the osculating orbit is precessing relative to the unperturbed orbit [see equations $(21)-(23)$ in Paper I] and the true anomaly time-evolution deviates from the first unperturbed term in equation $(\ref{true2})$. The additional terms come from the fact that, while time $t$ is measured from a fixed moment in time, $\nu$ is measured from the periapsis, which changes through both precessions $\dot{\omega}$ and $\dot{\Omega}$. For perturbations induced by mass-loss/transfer processes we study in this paper we proved in Section 3 in Paper I [see equation $(65)$ in Paper I] that $\dot{\Omega}=0$. Assuming in addition that in first order approximation the precession of the argument of periapsis $\omega$ due to mass-loss/transfer is small (i.e., $\dot{\omega} << 1$) then to first order we can compute the secular evolution of the orbital elements neglecting the second and third term in equation $(\ref{true2})$ and using instead the substitution
\begin{equation}
 d\nu=n\frac{(1+e\cos{\nu})^{2}}{(1-e^{2})^{3/2}}dt
\end{equation} 
when we integrate the equations $da/dt$ and $de/dt$ over $\nu$ from $0$ to $2\pi$. In order to incorporate higher-order effects in the perturbation equations we have to proceed one step further and include all the terms in equation $(\ref{true2})$ when applying the orbit-averaging process. In this paper, we derive first-order secular time-evolution equations where the general orbit-averaging rule is defined by the orbit-averaging integral
\begin{equation}
\left\langle\: (...) \:\right\rangle=\frac{(1-e^{2})^{3/2}}{2\pi}
\int_{0}^{2\pi}\: (...) \: \frac{d\nu}{(1+e\:cos{\nu})^{2}}\label{avint}
\end{equation}
where we made use only of the first term in equation $(\ref{true2})$ and the quantity to be orbit-averaged is depicted with the three dots and is generally phase-dependent. In most cases, this makes the above integral tedious and not easy to resolve into a simple analytical form. However, there are cases that lead to significant simplifications to the time-evolution equations for the secular orbital elements. Two commonly used simplifying approximations are the \emph{adiabatic approximation} and the assumption of a \emph{delta-function perturbation} that we will describe in detail in the following two subsections. In the next sections, we orbit-average the various phase-dependent time-evolution equations derived in Paper I under both of these approximations.
\subsection{Adiabatic approximation}

To proceed with orbit-averaging equations $(\ref{agen})$ and $(\ref{egen})$ we must make some assumptions about the perturbation timescale. If the perturbation timescale is much longer compared to the orbital timescale, the natural oscillations of the orbital elements can be smoothed out assuming the perturbation is nearly constant along the orbit. Keeping the perturbation an ``adiabatic variant" of the system, this assumption is called \emph{the adiabatic approximation} and the regime where the adiabatic approximation is valid is known as the \emph{adiabatic regime}. The system leaves this regime when the perturbation timescale becomes comparable to the orbital period. At this point adiabaticity is broken and the orbit-averaging method no longer applies \citep[e.g.,][]{2011MNRAS.417.2104V}.
\par For mass-transferring binary systems the adiabatic regime is valid by definition when the mass-loss rates involved in a problem are constants or under a reasonable approximation when the body's orbital period is much smaller than its mass-loss timescale. Mathematically, this means that the mass-loss is so slow that both the body's mass-loss rate and its mass can be assumed constant over one orbital period, i.e., we can take factors like $\dot{M}/M$ out of the averaging integral $(\ref{avint})$ assuming they are constant along the orbit. 
\subsection{Delta-function perturbation}

Under certain conditions, the effect of a perturbation on a two-body system can have its peak at specific orbital phases. In this case the perturbation force can be approximated like delta-function perturbation centered at specific points in the orbit.\par
In the case of mass-loss/transfer in eccentric orbits, the evolution of the mass-transfer rate has a Gaussian-like behavior, with a maximum value occurring at periastron \citep[e.g.,][]{2011ApJ...726...67L,2013A&A...556A...4D}. Thus, it is reasonable to assume that mass-loss will mostly occur near periastron, since this is the point in an eccentric orbit where the component stars are closest to each other. Because of this, it is logical to examine the case of delta-function mass-loss/transfer at periastron. In this case, we can assume a mass-loss rate $\dot{M}_{0}$ averaged over $2\pi$ such that
\begin{equation}
\dot{M}=\frac{\dot{M_{0}}}{2\pi}\:\:\delta(\nu).
\end{equation}
where $M$ is the mass of the mass-losing/accreting star and $\dot{M}$ the relevant mass-loss/accretion rate. The function $\delta(\nu)$ is the delta-function as a function of the true anomaly $\nu$ centered at the periastron position where $\nu=0$.
\section{Reference frame decomposition}

The time-evolution equations for the osculating orbital elements in their vector form [see equations $(18)$-$(23)$ in Paper I] are independent of the reference frame choice. Decomposing the perturbing force into its components in different coordinate systems is helpful and leads to different sets of equations for the  semi-major axis and the eccentricity in each system.\par
In Section 3 of Paper I, we presented the form of the phase-dependent time-evolution equations for the osculating orbital elements in three different reference systems [see Figure 1 here and Section 3 of Paper I for a detailed description of the reference systems]. 
\begin{figure}
  \centering
    \includegraphics[width=0.5\textwidth]{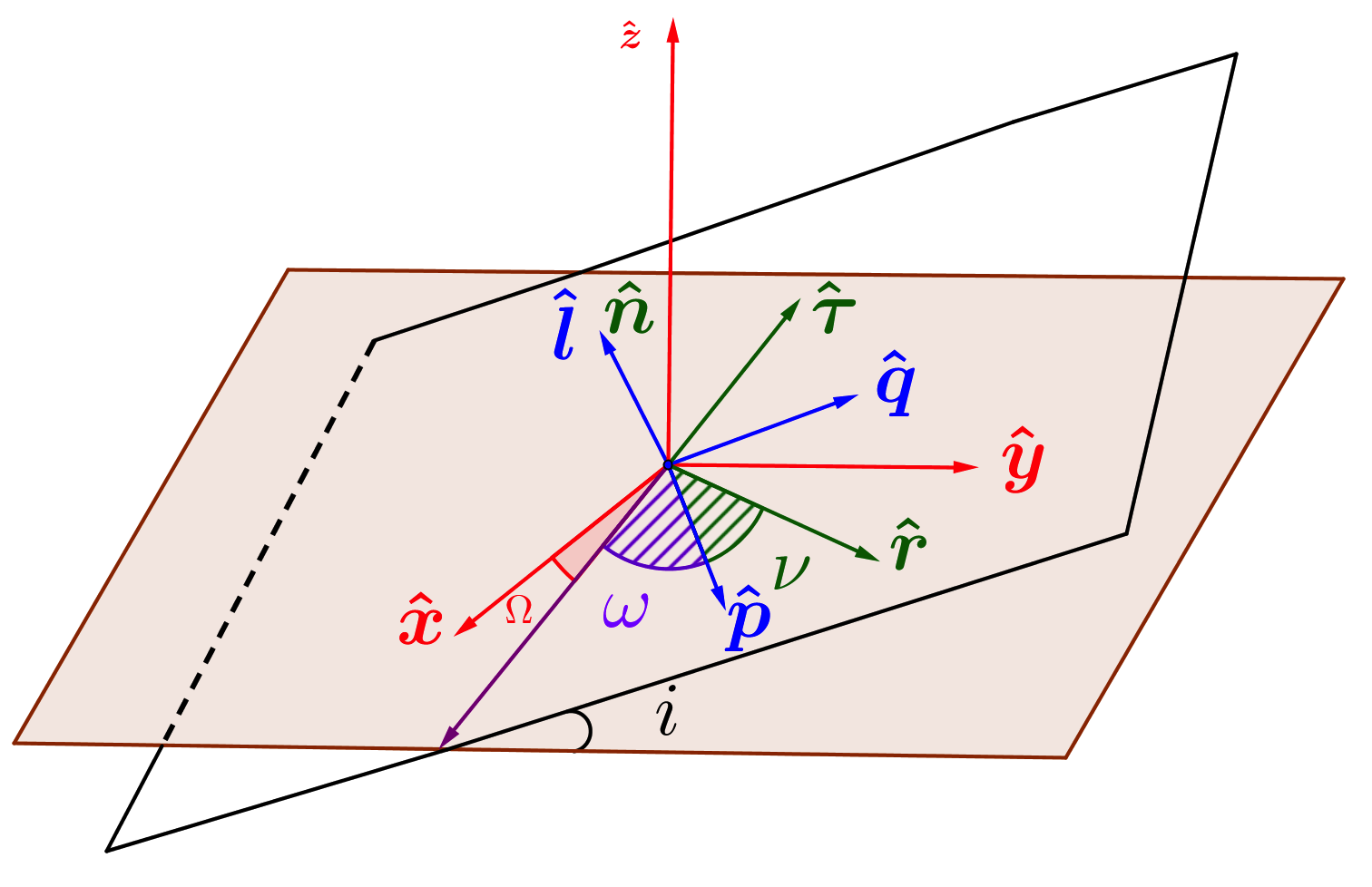}
      \caption{Definition of different reference frames. Solid shaped planes and angles refer to the inertial reference frame $(\bold{\hat{x}},\bold{\hat{y}},\bold{\hat{z}})$, while hatch shaped angles refer to the inclined by angle $i$ orbital plane. In the latter, two reference systems are defined, $(\bold{\hat{p}},\bold{\hat{q}},\bold{\hat{l}})$ and $(\bold{\hat{r}},\bold{\hat{\tau}},\bold{\hat{n}})$, rotated by angle $\nu$ relative to the former.}
\end{figure}
\par If we assume that the perturbation $\bold{A}$ is nearly constant (i.e., phase-independent) along the orbit, then under the adiabatic approximation equations $(44)$ and $(45)$ of Paper I orbit-averaged become
\begin{align}
\label{av1}\left\langle\frac{da}{dt}\right\rangle_{\bold{\hat{p}},\bold{\hat{q}},\bold{\hat{l}}}^{ad}&=0\\
\label{av2}\left\langle\frac{de}{dt}\right\rangle_{\bold{\hat{p}},\bold{\hat{q}},\bold{\hat{l}}}^{ad}&=-\frac{3}{2}\frac{(1-e^{2})^{1/2}}{na}A_{q}
\end{align}
while equations  $(46)$ and $(47)$ of Paper I yield
\begin{align}
\left\langle\frac{da}{dt}\right\rangle_{\bold{\hat{r}},\bold{\hat{\tau}},\bold{\hat{n}}}^{ad}&=\frac{2(1-e^{2})^{1/2}}{n}A_{T}\label{av3}\\
\left\langle\frac{de}{dt}\right\rangle_{\bold{\hat{r}},\bold{\hat{\tau}},\bold{\hat{n}}}^{ad}&=-\frac{3}{2}e\frac{(1-e^{2})^{1/2}}{na}A_{T}.\label{av4}
\end{align}
Equations $(\ref{av1})$ and $(\ref{av2})$ describe the secular evolution of the osculating semi-major axis and eccentricity for a phase-independent perturbation under the adiabatic approximation and in the reference system $K_{J}(\bold{\hat{p}},\bold{\hat{q}},\bold{\hat{l}})$ shown in Figure 1. Equations $(\ref{av3})$ and $(\ref{av4})$ describe the same but in the reference system $K_{R}(\bold{\hat{r}},\bold{\hat{\tau}},\bold{\hat{n}})$ shown also in Figure 1.
\section{Isotropic wind mass-loss}

In the case of isotropic wind mass-loss, there is spherical symmetry and thus the quantity $\bold{b}$ defined by equation $(\ref{generalb})$ is zero, i.e., $\bold{b}=0$. Since there is no mass-transferred we can set $h=0$. We assume that the mass lost does not remove any extra angular momentum other than the one lost due to the total mass change, i.e., $\zeta=0$. Substituting $\bold{b}=h=\zeta=0$ into equations $(\ref{agen})$ and $(\ref{egen})$ and performing the
averaging rule $(\ref{avint})$ under the adiabatic approximation ($\dot{M}/M$ constant along the orbit) leads to \emph{the secular time-evolution equations for the osculating semi-major axis and eccentricity in the adiabatic regime in the case of isotropic wind mass-loss} \citep[e.g.,][]{1963Icar....2..440H}
\begin{align}
\left\langle\frac{da}{dt}\right\rangle_{iso}^{ad}&=-\frac{\dot{M}}{M}a\label{isoada}\\
\left\langle \frac{de}{dt} \right\rangle_{iso}^{ad}&=0\label{isoade}
\end{align}
where $M$ is the total mass, $\dot{M}$ is the mass-loss rate and the subscript $iso$ stands for isotropic.
\par Equations $(\ref{isoada})$ and $(\ref{isoade})$ indicate that an isotropic wind mass-loss leads to an increase in the secular semi-major axis while keeping the secular eccentricity unchanged. In other words, under adiabatic isotropic wind mass-loss an initially circular orbit will remain on average circular while an initially eccentric remain will retain on average its initial eccentricity.
\par We mention here that \citet{2008AdSpR..42.1313G} considered the case in which only one body is losing mass isotropically. Following equation $(3)$ in their paper, they assume for this case a perturbation of the form $\bold{F}=-(\dot{M}_{1}/M_{1})\bold{\dot{r}}$ instead of  $\bold{F}=-(\dot{M}_{1}/M)\bold{\dot{r}}$. However, the latter is the correct perturbation for the system in the case only one component is losing mass in an isotropic way [see Section 4 in Paper I]. The former perturbation refers to case of anisotropic ejection of mass with an absolute velocity of zero [see Section 6 here]. This important difference leads to factor of $2$ as well as different denominator discrepancy in their equations $(31)$, where $\dot{a}=-2(\dot{M}_{1}/M_{1})a,\dot{e}=0$. The semi-major axis secular evolution equation is different compared to our equation $(\ref{isoada})$ due to both the different denominator and the factor of $2$ while the eccentricity secular evolution remains in both zero.
\par In Paper I we used the generalized Lagrange gauge $\bold{\Phi}=-\partial \Delta L/\partial \bold{\dot{r}}$ [see subsection 2.1.2 of Paper I] to derive the phase-dependent time-evolution equations for the \emph{contact orbital elements} [see equations $(73)$ and $(74)$ in Paper I]. If we orbit-average these equations under the adiabatic approximation we see that the secular evolution of these elements is zero which is different from what equations $(\ref{isoada})$ and $(\ref{isoade})$ indicate for the \emph{osculating} semi-major axis and eccentricity. However, we remind here that unlike the osculating orbital elements the contact orbital elements have no physical meaning but  are sometimes useful mathematical tools [see see subsection 2.1.2 of Paper I for a more detailed discussion]. In other words, this means that the fact that the contact orbital elements do not evolve secularly has no physical interpretation. 
\par Under the assumption of isotropic delta-function mass-loss at periastron, the orbit-averaging procedure $(\ref{avint})$ on equations $(\ref{agen})$ and $(\ref{egen})$ (after the substitution $\bold{b}=h=\zeta=0$) yields
\begin{align}
\left\langle\frac{da}{dt}\right\rangle_{iso}^{\delta}&=-\frac{\dot{M}}{M}(1-e^{2})^{1/2}a\label{isodeltaa}\\
\left\langle\frac{de}{dt}\right\rangle_{iso}^{\delta}&=
-\frac{\dot{M}}{M}(1-e^{2})^{1/2}(1-e).\label{isodeltae}
\end{align}
\par Equations $(\ref{isodeltaa})$ and $(\ref{isodeltae})$ indicate that isotropic wind mass-loss at periastron leads to a secular increases in the  both the semi-major axis and eccentricity.\par
We note here that throughout the whole process no assumption is made for low eccentricities, so the above equations are valid for any value of eccentricity. \par
Besides, equations $(\ref{isodeltaa})$ and $(\ref{isodeltae})$ show that the circular case is equivalent to the adiabatic isotropic wind mass-loss case [see equations $(\ref{isoada})$ and $(\ref{isoade})$]. This can be seen by plugging in $e=0$ to equations $(\ref{isodeltaa})$ and $(\ref{isodeltae})$.
\par For completeness, we mention here that for a perturbing force linearly proportional to the orbital velocity, i.e., $\bold{F} \propto \bold{\dot{r}}$ (e.g., adiabatic isotropic wind mass-loss studied here) the time-evolution of $\omega$ is $\propto \sin\nu$ which means that on average the change in $\omega$ is zero.
\section{Anisotropic wind mass-loss}

In Paper I, we derived the perturbing acceleration in the case of anisotropic wind mass-loss from the donor of mass $M_{1}(t)$ which is given by $\bold{A}_{aniso}=\bold{F}_{aniso}/M_{1}(t)$. For the sake of simplicity, we choose the reference system $K_{J}({\bold{\hat{p}},\bold{\hat{q}},\bold{\hat{l}}})$ [see Figure $1$]. Substituting the aforementioned perturbing acceleration $\bold{A}$ to the adiabatic equations $(\ref{av1})$ and $(\ref{av2})$ leads to the following \emph{secular time-evolution equations for the osculating semi-major axis and the eccentricity in the adiabatic regime for the case of anisotropic wind mass-loss}
\begin{align}
\label{av1aniso}\left\langle\frac{da}{dt}\right\rangle _{\bold{\hat{p}},\bold{\hat{q}},\bold{\hat{l}}}^{ad,aniso}&=0\\
\label{av2aniso}\left\langle\frac{de}{dt}\right\rangle _{\bold{\hat{p}},\bold{\hat{q}},\bold{\hat{l}}}^{ad,aniso}&=-\frac{3}{2}\frac{(1-e^{2})^{1/2}}{naM_{1}(t)}F_{aniso,q}.
\end{align}
where the components of the perturbing force $\bold{F}_{aniso}$ in the inertial  reference system $K_{I}({\bold{\hat{x}},\bold{\hat{y}},\bold{\hat{z}}})$ [see Figure $1$] are given by
\begin{scriptsize}
\begin{align}
F_{aniso,x}&=\frac{1}{4\pi}\int_{0}^{\pi}\int_{0}^{2\pi}J(\phi,\theta,t)u(\phi,\theta,t)d\phi \sin{\theta}\cos{\phi}d\theta\label{firstref}\\
 F_{aniso,y}&=\frac{1}{4\pi}\int_{0}^{\pi}\int_{0}^{2\pi}J(\phi,\theta,t)u(\phi,\theta,t)d\phi \sin{\theta}\sin{\phi}d\theta\\
 F_{aniso,z}&=\frac{1}{4\pi}\int_{0}^{\pi}\int_{0}^{2\pi}J(\phi,\theta,t)u(\phi,\theta,t)d\phi \cos{\theta}d\theta .\label{lastref}
\end{align}
\end{scriptsize}
and the transformation $K_{I}\rightarrow K_{J}$ is done as follows using the matrix $\bold{Q}$ defined by equations $(52)-(60)$ in Paper I
 \begin{equation}
 \begin{pmatrix}
  F_{aniso,p}\\
  F_{aniso,q}\\
  F_{aniso,l} \\
 \end{pmatrix}=
 \begin{pmatrix}
  Q_{11} & Q_{21}  & Q_{31} \\
  Q_{12}  & Q_{22} &  Q_{32} \\
  Q_{13} & Q_{23}  & Q_{33} 
 \end{pmatrix}
  \begin{pmatrix}
  F_{aniso,x}\\
  F_{aniso,y}\\
   F_{aniso,z}\\
 \end{pmatrix}.\label{matrix}
\end{equation}
It is obvious from equation $(\ref{av1aniso})$ that for a phase-independent perturbing force and under the adiabatic approximation, there is no secular evolution of the osculating semi-major axis due to phase-independent anisotropic wind mass-loss [see also \citet{2013MNRAS.435.2416V}]. On the other hand, according to equation $(\ref{av2aniso})$ a secular change will be induced to the eccentricity if $F_{aniso,q} \neq 0$. The validity of this relation depends on the form and structure of the anisotropic wind. In the next two subsections we discuss briefly and qualitatively the effect of different kind of anisotropic wind structures (including jets) on the secular eccentricity.
\subsection{Latitudinal and Longitudinal dependence}

Equation $(\ref{matrix})$ indicates that the only case in which we can have $F_{aniso,q}=0$ is when $F_{aniso,x}=F_{aniso,y}=F_{aniso,z}=0$. Under these conditions the secular eccentricity will remain unchanged. We will discuss qualitatively different types of anisotropic wind mass-loss, separating the cases of latitudinal and longitudinal dependence.\par
We assume that the mass-flux rate we can be decoupled in its space and time dependence as $J(\phi,\theta,t)=J_{0}\kappa(\theta)\xi(\phi)$, where $J_{0}=J(\theta=0,\phi=0,t)$.
\par For no longitudinal dependence we have $\xi(\phi)=1$. Equations $(\ref{firstref})-(\ref{lastref})$ show that in this case $F_{aniso,x}=F_{aniso,y}=0$ always.
\begin{enumerate}
\item The final form of $F_{aniso,z}$ depends on the functional form of the latitudinal dependence $\kappa(\theta)$. For an even function $\kappa(\theta)$ we have $F_{aniso,z}=0$. This means that if the wind has only latitudinal dependence and is \emph{symmetric about the equator}, there is no additional secular effect on the eccentricity.
\item Any asymmetry in the northern to southern hemisphere wind can be described by the relation $\kappa(\theta)\left\lbrace 0<\theta<\pi/2\right\rbrace=\eta\:\: \kappa(\theta)\left\lbrace\pi/2<\theta<\pi\right\rbrace$, where $\eta >1$ is a constant parameter that measures the asymmetry between the two hemispheres. This relation leads to $F_{aniso,q}\neq 0$ since $F_{aniso,z}\neq 0\Rightarrow F_{aniso,q}=Q_{32} F_{aniso,z}\neq 0$. Based on equation $(\ref{av2aniso})$ we get $\dot{e}\neq 0$ which means that the secular eccentricity changes. The actual sign and type of the secular change in eccentricity depends on the form of the wind latitudinal dependence $\kappa(\theta)$. 
\end{enumerate}
\par In the case of only longitudinal dependence, $\xi(\phi)\neq 0$ and $\kappa(\theta)=1$. Under this assumption, there is no way to simultaneously have $F_{aniso,x}=F_{aniso,y}=F_{aniso,z}=0$. More specifically, regardless of whether the mass is lost symmetrically from the western and eastern hemisphere or not, we have a secular change in the eccentricity that will depend on the form of the wind longitudinal dependence $\xi(\phi)$.
\par For general longitudinal and latitudinal dependence the secular eccentricity changes, while the form of the wind space dependence expressed through the functions $\kappa(\theta)$ and $\xi(\phi)$ determines the sign and type of the eccentricity secular time-evolution equation.
\subsection{Jets}

In thin bipolar jets, the mass-loss flux rate does not depend on the longitude $\phi$. Based on what we showed in the previous subsection 5.1 in this case we have $F_{aniso,x}=F_{aniso,y}=0$. Furthermore, it is reasonable to assume that in thin bipolar jets both the mass-loss flux rate and the wind velocity are constant and uniform. If the bipolar jets are also symmetric about the equator of the donor then we have $\dot{M}_{up}u_{up}=\dot{M}_{down}u_{down}$, where $up:\theta=0$ and $down:\theta=\pi$. This leads to $F_{aniso,z}=0$ as well and the secular eccentricity remains constant.\par
Asymmetric thin bipolar jets introduce a secular eccentricity evolution proportional to the quantity $\dot{M}_{down}u_{down}-\dot{M}_{up}u_{up}$.\par
A jet oriented in any other random direction in $(\phi,\theta)$, breaks the symmetry of the ejection and the secular eccentricity changes accordingly.
\section{Non-isotropic ejection/accretion in mass-transfer (reaction forces)}
\subsection{Conservative case $\dot{M}=\dot{J}_{orb}=0$}

In all the following sections we choose to decompose our perturbations in the reference system $K_{R}$ [see Figure 1]. \par When all the mass ejected from the donor is accreted by the companion, there is no change in the total mass and orbital angular momentum of the system. For the so-called conservative case, this leads to $\dot{M}=0$ in equations $(\ref{agen})$ and $(\ref{egen})$.
\par \emph{The secular time-evolution equations $(\ref{agen})$ and $(\ref{egen})$ for the case of conservative mass-transfer} (i.e., $\dot{M}=0$) in the \emph{adiabatic regime} for extended bodies lead to
\begin{align}
\begin{split}
\left\langle\frac{da}{dt}\right\rangle_{con}^{ad}&=
\frac{2}{n\sqrt{1-e^{2}}}\left[e\left\langle b_{r}\sin{\nu}\right\rangle \right. \\
&\left. +\left\langle b_{\tau}(1+e\cos{\nu})\right\rangle\right]-2ha
\end{split}\label{conss1}\\
\begin{split}
\left\langle\frac{de}{dt}\right\rangle_{con}^{ad}&=\frac{(1-e^{2})^{1/2}}{na}\left[ \left\langle b_{r}\sin{\nu}\right\rangle \right. \\
&\left. +\left\langle b_{\tau}\frac{2\cos{\nu}+e(1+\cos^{2}\nu)}{1+e\cos{\nu}}\right\rangle\right]
\end{split}\label{conss2}
\end{align}
while for a \emph{delta-function mass-loss/transfer at periastron} we have
\begin{align}
\begin{split}
 \left\langle\frac{da}{dt}\right\rangle_{con}^{\delta}&=\frac{2}{n\sqrt{1-e^{2}}}\left[e\left\langle b_{r}\sin{\nu}\right\rangle \right.\\
&\left. +\left\langle b_{\tau}(1+e\cos{\nu})\right\rangle\right]
-2ha(1-e^{2})^{1/2}
\end{split}\label{conss3}\\
\begin{split}
\left\langle\frac{de}{dt}\right\rangle_{con}^{\delta}&=\frac{(1-e^{2})^{1/2}}{na}\left[\left\langle b_{r}\sin{\nu}\right\rangle\right. \\
&\left. +\left\langle b_{\tau}\frac{2\cos{\nu}+e(1+\cos^{2}\nu)}{1+e\cos{\nu}}\right\rangle \right] \\
&-2h(1-e^{2})^{1/2}(1-e).
\end{split}\label{conss4}
\end{align}
\par We remind here that first terms in square brackets on the RHS of equations $(\ref{conss1})-(\ref{conss4})$ depend through the quantity $\bold{b}$ defined in equation $(\ref{generalb})$ on the ejection and accretion points and associated velocities. The second terms on the RHS emerge from the mass-transferred from the donor to the accretor and are proportional through the quantity $h$ defined in equation $(\ref{h2})$ to the mass-loss/transfer rate $\dot{M}_{2}=-\dot{M}_{1}$. 
\par In Paper I we used the generalized Lagrange gauge $\bold{\Phi}=-\partial \Delta L/\partial \bold{\dot{r}}$ [see section 2.1.2 of Paper I] to derive the phase-dependent time-evolution equations for the contact elements [see equations $(105)$ and $(106)$ in Paper I]. Either under the adiabatic approximation or the delta-function mass-loss/transfer at periastron, equations $(105)$ and $(106)$ in Paper I give for the secular contact semi-major axis and eccentricity 
\begin{align}
\begin{split}
\left\langle\frac{da_{con}}{dt}\right\rangle_{con}^{ad,\delta}&=\frac{2}{n\sqrt{1-e^{2}}}\left[e\left\langle b_{r}\sin{\nu}\right\rangle
\right.\\ &\left. +\left\langle b_{\tau}(1+e\cos{\nu})\right\rangle \right]
\end{split}\label{contact1}\\
\begin{split}
\left\langle\frac{de_{con}}{dt}\right\rangle_{con}^{ad,\delta}&=\frac{(1-e^{2})^{1/2}}{na}\left[\left\langle b_{r}\sin{\nu}\right\rangle
\right. \\ &\left. +\left\langle b_{\tau}\frac{2\cos{\nu}+e(1+\cos^{2}\nu)}{1+e\cos{\nu}}\right\rangle \right].\label{contact2}
\end{split}
\end{align}
\par From equations $(\ref{contact2})$ and $(\ref{conss2})$ we see that the secular evolution for contact and osculating eccentricity under the adiabatic approximation is the same while from $(\ref{contact1})$ and $(\ref{conss1})$ the secular evolution of contact and osculating semi-major axis is different.
\par Under delta-function mass-loss/transfer at periastron equations $(\ref{contact1})$ and $(\ref{conss3})$ as well as $(\ref{contact2})$ and $(\ref{conss4})$ show that both the secular evolution of contact and osculating eccentricity and semi-major axis are different. We remind here again that the contact orbital elements have no physical meaning but still remain useful mathematical tools. 
\subsubsection{Conservative delta-function RLOF}

In the case of Roche-Lobe-Overflow (RLOF) the mass is lost from the donor through the Lagrangian point $L_{1}$. This means that for RLOF conservative mass-transfer we have $\bold{r}_{A_{1}}=\bold{r}_{L_{1},P},\bold{r}_{A_{2}}=\bold{R}_{acc}$, where $L_{1},P$ is the location of the Lagrangian point at periastron with respect to the center of mass of the donor and $\bold{R}_{acc}$ is the radius vector of the accretor's surface with respect to its center of mass.\par
\citet{2007ApJ...667.1170S} assumed RLOF conservative delta-function mass-loss/transfer at periastron so that $\dot{M}_{2}=-\dot{M}_{1}=-(\dot{M}_{0}/2\pi)\delta(\nu)$. Although not explicitly stated but implied by their equation $(35)$ they assumed so-called stationary particles which have zero absolute velocities, i.e., $(\ref{generalb})$ $\bold{w}_{1}=\bold{w}_{2}=0$. Under this assumption equation $(\ref{generalb})$ becomes
\begin{align}
\nonumber \bold{b}=&\dot{M}_{2}\left(
+\frac{\bold{\omega}_{orb}\times \bold{r}_{A_{2}}}{M_{2}}
+\frac{\bold{\omega}_{orb}\times \bold{r}_{A_{1}}}{M_{1}}
\right)\\&+\ddot{M}_{2}\left(\frac{\bold{r}_{A_{2}}}{M_{2}}
+\frac{\bold{r}_{A_{1}}}{M_{1}}\right).\label{betarlo}
\end{align}
\par From equations $(\ref{conss3})$ and $(\ref{conss4})$  we see that for delta-function mass-loss/transfer since $\sin{\nu}=0$ at periastron and both $\bold{r}_{L_{1},P}$ and $\bold{R}_{acc}$ are not phase-dependent quantities, the only term that depends on $\bold{b}$ and survives in equations $(\ref{conss3})$ and $(\ref{conss4})$ is the one including the tangential component of $\bold{b}$ i.e., $b_{\tau}$.
\par Substituting equations $(\ref{h2})$ and $(\ref{betarlo})$ into $(\ref{cons3})$ and $(\ref{cons4})$ and after some algebra we are led to the following \emph{secular time-evolution equations for the osculating semi-major axis and eccentricity in the case of conservative RLOF} \citep{2007ApJ...667.1170S} 
\begin{align}
\nonumber \left\langle\frac{da}{dt}\right\rangle_{con}^{\delta, \begin{tiny} RL\end{tiny}}&=
-2a\frac{\dot{M}_{2}}{M_{1}}\frac{1}{(1-e^{2})^{1/2}}\left[
qe\frac{r_{A_{2}}}{a}\cos{\phi_{p}}+e\frac{r_{A_{1}}}{a}\right]\\
&-2\dot{M}_{2}\left(\frac{1}{M_{2}}-\frac{1}{M_{1}}\right)a(1-e^{2})^{1/2}
\label{deltarloa}\\
\nonumber \left\langle\frac{de}{dt}\right\rangle_{con}^{\delta,RL}&=
-\frac{\dot{M}_{2}}{M_{1}}(1-e^{2})^{1/2}\left[
q\frac{r_{A_{2}}}{a}\cos{\phi_{p}}+\frac{r_{A_{1}}}{a}\right]\\
&-2\dot{M}_{2}\left(\frac{1}{M_{2}}-\frac{1}{M_{1}}\right)(1-e^{2})^{1/2}(1-e)\label{deltarloe}
\end{align}
where $q=M_{1}/M_{2}$ is the mass ratio and $\phi_{p}$ is the angle between the unit vector $\bold{\hat{r}}$ and the vector from the center of mass of the accretor to the accretion point $A_{2}$ at periastron. \par
Equations $(\ref{deltarloa})$ and $(\ref{deltarloe})$ are the same as equations $(39)$ and $(40)$ of \citet{2007ApJ...667.1170S}. \citet{2007ApJ...667.1170S} compared
the timescale for the evolution of the eccentricity due to
tides and RLOF and found that, in contrast to tides which always
act to circularize the orbit, mass-transfer may either increase
or decrease the eccentricity, over timescales ranging from
a few Myr to a Hubble time. Furthermore, the timescale over
which mass-transfer acts to increase the eccentricity may be
shorter than the tidal timescale which acts to decrease it.\par
For completeness, we note here that equations $(\ref{deltarloa})$ and $(\ref{deltarloe})$ reduce to the point-mass for stationary particles equations $(\ref{mya})$ and $(\ref{mye})$ described in section 7.4 by setting $\bold{r}_{A_{1}}=\bold{r}_{A_{2}}=0$. Thus, the first two terms in the RHS of equations $(\ref{deltarloa})$ and $(\ref{deltarloe})$ emerge from the fact that we took into account the physical extent of the ejection and accretion point, which introduces additional perturbations in our problem.
These perturbations can be seen by looking into terms dependent on $\bold{r}_{A_{1}},\bold{r}_{A_{2}}$ in equation $(\ref{generalb})$.

\subsection{Non-conservative case, $\dot{M},\dot{J}_{orb}\neq 0$}

When a fraction of the mass ejected from the donor is accreted by the companion and the residual mass is lost from the system, there is change in the total mass and orbital angular momentum of the system. For this so-called non-conservative case, $\dot{M}\neq 0$ and in the most general case $\zeta \neq 0$ too in equations $(\ref{agen})$ and $(\ref{egen})$. We mention here that from now on in all cases with $\zeta \neq 0$ when we perform the orbit-average in the adiabatic regime we assume that the parameter $\zeta$ is changing adiabatically so that we can take it out from the orbit-averaging integral $(\ref{avint})$.
\par \emph{The secular time-evolution equations $(\ref{agen})$ and $(\ref{egen})$ for the case of non-conservative mass-transfer} (i.e., $\dot{M}=0$) in the \emph{adiabatic regime} for extended bodies lead to
{\setlength{\abovedisplayskip}{6pt}
\setlength{\belowdisplayskip}{0pt}
\begin{align}
\begin{split}
\left\langle\frac{da}{dt}\right\rangle_{non-con}^{ad}=&
\frac{2}{n\sqrt{1-e^{2}}}\left[e\left\langle b_{r}\sin{\nu}\right\rangle \right. \\
&\left. +\left\langle b_{\tau}(1+e\cos{\nu})\right\rangle\right]-2ha
\\&-(\zeta+1)\frac{\dot{M}}{M}
\end{split}\label{cons1}\\
\begin{split}
\left\langle\frac{de}{dt}\right\rangle_{non-con}^{ad}=&\frac{(1-e^{2})^{1/2}}{na}\left[ \left\langle b_{r}\sin{\nu}\right\rangle \right. \\
&\left. +\left\langle b_{\tau}\frac{2\cos{\nu}+e(1+\cos^{2}\nu)}{1+e\cos{\nu}}\right\rangle\right]
\end{split}\label{cons2}
\end{align}}
while for \emph{delta-function mass-transfer at periastron} we have
\begin{align}
\begin{split}
 \left\langle\frac{da}{dt}\right\rangle_{non-con}^{\delta}=&\frac{2}{n\sqrt{1-e^{2}}}\left[e\left\langle b_{r}\sin{\nu}\right\rangle \right.\\
&\left. +\left\langle b_{\tau}(1+e\cos{\nu})\right\rangle\right]
-2ha(1-e^{2})^{1/2}
\\&-(\zeta+1)\frac{\dot{M}}{M}(1-e^{2})^{1/2}
\end{split}\label{cons3}\\
\begin{split}
\left\langle\frac{de}{dt}\right\rangle_{non-con}^{\delta}=&\frac{(1-e^{2})^{1/2}}{na}\left[\left\langle b_{r}\sin{\nu}\right\rangle\right. \\
&\left. +\left\langle b_{\tau}\frac{2\cos{\nu}+e(1+\cos^{2}\nu)}{1+e\cos{\nu}}\right\rangle \right] \\
&-2h(1-e^{2})^{1/2}(1-e)
\\&-(\zeta+1)\frac{\dot{M}}{M}(1-e^{2})^{1/2}(1-e).
\end{split}\label{cons4}
\end{align}
Comparing equations $(\ref{conss1})-(\ref{conss4})$ with $(\ref{cons1})-(\ref{cons4})$ we see an additional third term arising on the RHS of the above equations. This term emerges from the non-conservative nature of the mass-transfer which carries away from the system both mass and orbital angular momentum. The first component of this term is proportional to the parameter $\zeta$ describing the loss in the system's orbital angular momentum while the second component describes the change in the total system mass and is proportional to total mass-loss rate $\dot{M}$.
\par The additional effect induced by an extra angular momentum removal from the system can be seen in equations $(\ref{cons1})$ and $(\ref{cons2})$ or $(\ref{cons3})$ and $(\ref{cons4})$. Assuming an adiabatically changing $\zeta \neq 0$ this additional removal of angular momentum enhances the increase rate of the secular semi-major axis and has no effect in the secular eccentricity. Within the same considerations the limiting case $\zeta \rightarrow 1$ does not induce any secular change in eccentricity but speeds up to the maximum level possible from this effect the secular expansion of the orbit. In the case of delta-function mass-transfer at periastron $\zeta \neq 0$  amplifies the increase rate of the semi-major axis while depending on the system under study speeds up the increase or decrease of eccentricity. Similarly to the adiabatic case, the limiting case $\zeta \rightarrow 1$ maximizes the aforementioned effects.
\subsubsection{Non-Conservative delta-function RLOF}

For non-conservative delta-function RLOF we assume that only a fraction $\gamma$ of the lost mass is accreted such that $\dot{M}_{2}=-\gamma\dot{M}_{1}$. Following a similar procedure with the one described in subsection 6.1.1, equations $(\ref{agen})$ and $(\ref{egen})$ give the following for the \emph{secular time-evolution equations for the osculating semi-major axis and eccentricity in the case of non-conservative RLOF}
\begin{align}
\begin{split}
\left\langle\frac{da}{dt}\right\rangle_{non-con}^{\delta,RL}&=
-2a\frac{\dot{M}_{2}}{M_{1}}\frac{1}{(1-e^{2})^{1/2}}\left[
\gamma qe\frac{r_{A_{2}}}{a}\cos{\phi_{p}}\right.\\
&\left.+\gamma e\frac{r_{A_{1}}}{a}\right]
-2\frac{\dot{M}_{2}}{M_{1}}a(1-e^{2})^{1/2}\left[\gamma (q-1)
\right.\\
&\left. +(1-\gamma)\left(\zeta+\frac{1}{2}\right)\frac{q}{1+q}\right]
\end{split} \label{rlononcon1}
\\
\begin{split}
\left\langle\frac{de}{dt}\right\rangle_{non-con}^{\delta,RL}&=
-\frac{\dot{M}_{2}}{M_{1}}(1-e^{2})^{1/2}\left[
\gamma q\frac{r_{A_{2}}}{a}\cos{\phi_{p}}\right.\\
&\left.+\gamma \frac{r_{A_{1}}}{a}\right]
-2\frac{\dot{M}_{2}}{M_{1}}(1-e^{2})^{1/2}(1-e)\left[\gamma (q-1)
\right.\\
&\left. +(1-\gamma)\left(\zeta+\frac{1}{2}\right)\frac{q}{1+q}\right].
\end{split}\label{rlononcon2}
\end{align}
\par Equations $(\ref{rlononcon1})$ and $(\ref{rlononcon1})$ do not change qualitatively the results for the mass-transfer timescale and the mass-transfer effect on the secular semi-major axis and eccentricity we mentioned in subsection 6.1.1 for the case of conservative mass-transfer.\par
These equations are also derived in \citet{2009ApJ...702.1387S}. However, in that study it was assumed that mass is lost only from the accretor and not from the system as a whole as we assume here and is described in detail in section 6.2 in Paper I. This difference leads to the factor $\gamma (q-1)$ in our equations instead of $(\gamma q-1)$ in equations $(18)$ in $(19)$ in \citet{2009ApJ...702.1387S}. The same reason also leads to the missing $\gamma$ factor in the first term in the RHS of equation $(18)$ and the second term in the RHS of equation $(19)$ in \citet{2009ApJ...702.1387S}.

\section{Point-mass approximation}

For the case of both binary components treated as point-masses, we have derived the form of the perturbing force [see equations $(119)$-$(122)$ in Paper I] as well as the phase-dependent time-evolution equations $(\ref{agenpoint})$ and $(\ref{egenpoint})$ for the semi-major axis and the eccentricity, respectively. In the following sections, we apply these equations in some commonly used characteristic examples of mass-loss and mass-transfer under the point-mass approximation. Furthermore, we compare our equations with equations derived other studies that have considered applications similar to the ones presented here.

\subsection{Radial ejection from the donor with no accretion}

We assume that the mass ejected from the donor is always in the radial direction pointing towards the accretor while none of this matter is accreted. In this case, we have $\dot{M}_{2}=0,\dot{M}=\dot{M}_{1},\zeta=0$ and $\bold{w}_{1}-\bold{v}_{1}=V_{r}\bold{\hat{r}}$. The quantity $\bold{c}$ defined by equation $(\ref{point1})$ then becomes 
\begin{equation}
 \bold{c}_{radial}=-\frac{\dot{M_{1}}}{M_{1}}\left( \bold{w}_{1}-\bold{v}_{1}
 \right)\label{c}
\end{equation}
from which in this case we have $c_{r}=-\frac{\dot{M_{1}}}{M_{1}}V_{r}$ and $c_{\tau}=0$ in equations $(\ref{agenpoint})$ and $(\ref{egenpoint})$.\\
We introduce a parameter $\lambda _{r}$ which is a measure of the ejection velocity as a fraction of the orbital velocity at periastron and is defined by
\begin{equation}
V_{r}=\lambda _{r} \frac{2\pi a}{P}.\label{lambda}
\end{equation}
\par The \emph{secular point-mass time-evolution equations $(\ref{agenpoint})$ and $(\ref{egenpoint})$ for the semi-major axis and the eccentricity in the adiabatic regime for radial ejection with no accretion} then become
\begin{align}
\left\langle\frac{da}{dt}\right\rangle_{r}^{ad}&=\
-\frac{\dot{M}_{1}}{M}a\label{arad}\\
\left\langle\frac{de}{dt}\right\rangle_{r}^{ad}&=0.\label{erad}
\end{align}
\par From equations $(\ref{arad})$ and $(\ref{erad})$ we see that a radial ejection from the donor with no accretion is equivalent to the case of isotropic wind mass-loss, which was studied in Section 4 and is governed by equations $(\ref{isoada})$ and $ (\ref{isoade})$. Equations $(\ref{arad})$ and $(\ref{erad})$ are consistent with the work of \citet{1956AJ.....61...49H}.

\subsection{Tangential ejection from donor with no accretion}

In this case the mass is ejected in the tangential direction while the assumption of no accretion remains the same as in the section 7.1. This means that $\dot{M}_{2}=0,\dot{M}=\dot{M}_{1},\zeta=0$ and $\bold{w}_{1}-\bold{v}_{1}=\pm V_{\tau}\bold{\hat{\tau}}$. In the last relation, for $V_{\tau}>0$, the plus sign refers to ejection in the direction opposite to the direction of motion of the body. We can then write the quantity $\bold{c}$ defined in equation $(\ref{point1})$ as
\begin{equation}
 \bold{c}_{tangential}=-\frac{\dot{M_{1}}}{M_{1}}\left( \bold{w}_{1}-\bold{v}_{1}
 \right)
\end{equation}
from which in this case we have $c_{r}=0$ and $c_{\tau}=\mp \frac{\dot{M_{1}}}{M_{1}} V_{\tau}$ in equations $(\ref{agenpoint})$ and $(\ref{egenpoint})$.
\par We introduce here the parameter $\lambda_{\tau} >0$ defined in the same way as in equation $(\ref{lambda})$ by
\begin{equation}
V_{r}=\lambda _{\tau} \frac{2\pi a}{P}.\label{lambda2}
\end{equation}
\par The \emph{secular point-mass time-evolution equations $(\ref{agenpoint})$ and $(\ref{egenpoint})$ for the semi-major axis and the eccentricity in the adiabatic regime for tangential ejection with no accretion} then become
\begin{align}
\left\langle\frac{da}{dt}\right\rangle_{t}^{ad}&=
-\frac{\dot{M}_{1}}{M}a\mp 2\lambda_{\tau} (1-e^{2})^{1/2}\frac{\dot{M}_{1}}{M_{1}}a\label{rear0}\\
\nonumber \left\langle\frac{de}{dt}\right\rangle_{t}^{ad}&=\mp \lambda_{\tau} 
\frac{(1-e^{2})^{3/2}}{e}\frac{\dot{M}_{1}}{M_{1}}\\
&\pm \lambda_{\tau} 
\left(\frac{e^{2}}{2}+1\right)\frac{(1-e^{2})^{1/2}}{e}\frac{\dot{M}_{1}}{M_{1}}\label{rear00}
\end{align}
where we remind here that the plus sign refers to ejection in the direction of motion of the body (front side) and the minus sign to the opposite (rear side). Equation $(\ref{rear00})$ seems to have a singularity for circular orbits ($e=0$). However, as we describe below equation $(\ref{rear00})$ reduces to simpler forms with no singularity for either rear and frontal ejection.
\par Specifically, for rear ejection, tangential but opposite to the orbital motion, we have the minus sign and equations $(\ref{rear0})$ and $(\ref{rear00})$ simplify to
\begin{align}
\left\langle\frac{da}{dt}\right\rangle_{t}^{ad}&=
-\frac{\dot{M}_{1}}{M}a- 2\lambda_{\tau} (1-e^{2})^{1/2}\frac{\dot{M}_{1}}{M_{1}}a\label{rear1}\\
\left\langle\frac{de}{dt}\right\rangle_{t}^{ad}&=\frac{3}{2} \lambda_{\tau} e(1-e^{2})^{1/2}.\label{rear2}
\end{align}
\par Equations $(\ref{rear1})$ and $(\ref{rear2})$ indicate that \emph{an ejection of mass opposite to the motion of the body direction leads to an increase in the secular semi-major axis and a decrease in the secular eccentricity of the orbit.}
\par For frontal ejection, tangential but along the orbital motion, we have the opposite result since in this case
\begin{align}
\left\langle\frac{da}{dt}\right\rangle_{t}^{ad}&=
-\frac{\dot{M}_{1}}{M}a+2\lambda_{\tau} (1-e^{2})^{1/2}\frac{\dot{M}_{1}}{M_{1}}a\label{front1}\\
\left\langle\frac{de}{dt}\right\rangle_{t}^{ad}&=-\frac{3}{2} \lambda_{\tau} e(1-e^{2})^{1/2}\label{front2}
\end{align}
which shows that \emph{an ejection in the same to the motion of the body direction leads to a secular semi-major axis decrease and a secular eccentricity increase}. Equations $(\ref{rear1})-(\ref{front2})$ for the circular case of $e=0$ predict a zero secular evolution for the eccentricity which means that an initially-circular orbit will remain circular under the assumption of only tangential adiabatic mass-loss with no accretion in the case of point masses. We note here that both quantitatively and qualitatively we agree with \citet{1956AJ.....61...49H}, who used a different formalism to derive equations similar to our equations $(\ref{rear1})-(\ref{front2})$. 
\par It is interesting to see at this point what happens in the rear ejection case for delta-function mass-loss at periastron. In this case we have
\begin{align}
\left\langle\frac{da}{dt}\right\rangle_{t}^{\delta}&=
-\frac{\dot{M}_{1}}{M}(1-e^{2})^{1/2}-\frac{\lambda_{\tau}}{\pi} (1-e)\frac{\dot{M}_{1}}{M_{1}}a\\
\left\langle\frac{de}{dt}\right\rangle_{t}^{\delta}&=-\frac{\lambda_{\tau}}{\pi}(1-e)^{2}\frac{\dot{M}_{1}}{M_{1}}
\end{align}
which depicts that \emph{both the secular semi-major axis and eccentricity increase for rear delta-function mass-loss at periastron}. If the same ejection is to take place at apastron, one can prove that this will lead to an increase in the secular semi-major axis and a decrease to the secular eccentricity.\par
Since the case of frontal ejection differs only by a sign compared to the rear ejection case, \emph{frontal delta-function mass-loss at periastron will lead to a semi-major axis and eccentricity secular decrease}, while delta-function mass-loss at apastron will decrease the secular semi-major axis and increase the secular eccentricity.

\subsection{Isotropic/Radial wind mass-loss from the donor, radial partial wind accretion and tangential mass-loss from the accretor}

We have shown in Section 7.1 that a radial ejection of mass is in the adiabatic regime secularly equivalent to the isotropic wind mass-loss case studied in Section 4. Following the same logic the same is true for any radially accreted matter.\par
For our case here we assume the donor is losing mass either isotropically in a wind or with a radial ejection pointing always towards the accretor. Part of this mass is accreted radially to the companion and the rest of it is tangentially lost from it with a constant absolute velocity $\bold{V}$.  What introduces perturbing forces in the scenario described above is the tangential mass-loss and we assume for it to always happen with a velocity opposite to the tangential motion of the accreting body. We note here that under these assumptions, the relative velocity of the tangentially lost mass to the center of mass of the companion is phase-dependent.\par
Based on the above assumptions, we have $\dot{M}_{2}=-\gamma \dot{M}_{1},\dot{M}=(1-\gamma)\dot{M}_{1}$ and $\bold{w}_{2}= V\bold{\hat{\tau}}$.  Because the quantity $\bold{c}$ defined by equation $(\ref{point1})$ is now phase-dependent, the tangential wind mass-loss introduces an effective $\zeta=\zeta_{eff}$. We note here that as we mentioned in Section 7.1 a radial ejection is in adiabatic regime secularly equivalent to the isotropic wind mass-loss and does not introduce any reaction force. Hence, the perturbing force defined by equation $(119)$ in Paper I, is only due to the tangential mass-loss and can be written as
\begin{align}
\bold{F}_{tan-loss}&=\frac{\dot{M_{2}}}{M_{2}}\left( \bold{w}_{2}-\bold{v}_{2}
 \right)-\frac{1}{2}\frac{\dot{M}}{M}\dot{\bold{r}}\\
 &=\frac{\dot{M_{2}}}{M_{2}}V\bold{\hat{\tau}}-\left(\frac{1}{2}\frac{\dot{M}}{M}+\frac{\dot{M_{2}}}{M_{2}}\right)\dot{\bold{r}}.\label{newway}
 \end{align}
From equation $(\ref{newway})$ we have for the value of $\bold{c}$ and $\zeta_{eff}$
 \begin{align}
 \bold{c}&=\frac{\dot{M_{2}}}{M_{2}} V\bold{\hat{\tau}}\label{newc}\\
 \zeta_{eff}&=\frac{2\gamma}{\gamma-1}\left(\frac{M_{1}}{M_{2}}+1\right).\label{zetaeff}
\end{align}
Based on equations  $(\ref{newc})$ and $(\ref{zetaeff})$ we have in equations $(\ref{agenpoint})$ and $(\ref{egenpoint})$ that $c_{r}=0$, $c_{\tau}=\frac{\dot{M_{2}}}{M_{2}} V$ and $\zeta=\zeta_{eff}$.
\par Similarly to the previous subsections, we introduce here the parameter $\lambda_{a}$ such that  
\begin{equation}
\frac{1-\gamma}{\gamma}V=\lambda_{a} \frac{2\pi a}{P}.
\end{equation}
\par We find that apart from the known secular equations for isotropic wind mass-loss/radial ejection in the adiabatic regime
\begin{align}
\left\langle\frac{da}{dt}\right\rangle^{ad}&=\
-\frac{\dot{M}_{1}}{M}a\label{againisoa}\\
\left\langle\frac{de}{dt}\right\rangle^{ad}&=0\label{againisoe}
\end{align}
the secular time-evolution equations $(\ref{agenpoint})$ and $(\ref{egenpoint})$ in the case studied here become
\begin{align}
\left\langle\frac{da}{dt}\right\rangle^{ad}&=
-\frac{1-\gamma}{\gamma}\frac{\dot{M}_{2}}{M}a- 2\lambda (1-e^{2})^{1/2}\frac{\dot{M}_{2}}{M_{2}}a\label{tan1}\\
\begin{split}
\left\langle\frac{de}{dt}\right\rangle^{ad}&=\left\lbrace \frac{3}{2}\lambda e(1-e^{2})^{1/2}\right.\\ &\left.
\hspace{1cm}+\:\:\rho\frac{(1-e^{2})^{3/2}}{e}\right\rbrace\frac{\dot{M}_{2}}{M_{2}}\label{tan2}
\end{split}
\end{align}
where $\rho=(1-e^{2})^{-1/2}-1$. In this case, equations $(\ref{tan1})$
and $(\ref{tan2})$ indicate that the secular semi-major axis decreases while the secular eccentricity increases. We note here that equations $(\ref{againisoa})$
and $(\ref{againisoe})$ are included in equations $(\ref{tan1})$
and $(\ref{tan2})$.\par
We mention here that we agree with the results by \citet{1988A&A...205..155B}. In that study \citet{1988A&A...205..155B} based on the work of \citet{1956AJ.....61...49H} made some corrections and re-derived equations $(\ref{tan1})$ and $(\ref{tan2})$. \citet{2008A&A...480..797B} derived similar equations to our equations $(\ref{tan1})$ and $(\ref{tan2})$ by making specific assumptions for the radial and tangential components of the wind velocity when the latter is accreted. As it can be seen from the general equations $(\ref{agen}) $ and $(\ref{egen})$, the final expressions for the time-evolution of $a$ and $e$ depend strongly on the specific assumptions made about wind velocity when it is ejected or is accreted. Thus, comparison between the time-evolution equations for the semi-major axis and eccentricity between previous work is strongly driven by the different assumptions made separately in each of these works.

\subsection{Conservative mass-transfer for stationary particles}

In many previous work, the ejected or accreted mass particles were assumed to not have an absolute velocity on their own, \citep[e.g.,][]{2000A&A...357..557S,2006epbm.book.....E,2007ApJ...667.1170S}. This assumption of the so-called \emph{stationary point particles} leads to the additional relation $\bold{w}_{1}=\bold{w}_{2}=0$, so that from equation $(\ref{generalb})$ we have for this case $\bold{b}=0$.\par
Under these assumptions, equations $(\ref{agen})$ and $(\ref{egen})$ lead to the following \emph{secular time-evolution equations for the semi-major axis and eccentricity in the adiabatic regime for stationary point particles and conservative mass-transfer}
\begin{align}
\left\langle\frac{da}{dt}\right\rangle_{con}^{ad}&=-2\dot{M}_{2}\left(\frac{1}{M_{2}}-\frac{1}{M_{1}}\right)a\label{adstation1}\\
\left\langle\frac{de}{dt}\right\rangle_{con}^{ad}&=0\label{adstation2}
\end{align}
while for \emph{delta-function conservative mass-transfer at periastron} we have
\begin{align}
 \left\langle\frac{da}{dt}\right\rangle_{con}^{\delta}&=
-2\dot{M}_{2}\left(\frac{1}{M_{2}}-\frac{1}{M_{1}}\right)a(1-e^{2})^{1/2}\label{mya}\\
\left\langle\frac{de}{dt}\right\rangle_{con}^{\delta}&=-2\dot{M}_{2}\left(\frac{1}{M_{2}}-\frac{1}{M_{1}}\right)(1-e^{2})^{1/2}(1-e).\label{mye}
\end{align}
\par Equations $(\ref{adstation1})-(\ref{mye})$ show that, in the case of stationary particles, the effect of mass-transfer on the secular semi-major axis and eccentricity depends on the mass ratio $q$. In the adiabatic regime, the secular eccentricity does not change while the secular semi-major axis decreases for $q>1$ and increases for $q<1$. In the case of delta-function mass-transfer at periastron, both the secular semi-major axis and the eccentricity 
decrease for $q>1$ and increase for $q<1$. \par
We mention here that the same perturbation for conservative mass-transfer and stationary particles, which is equal to $-\dot{M}_{2}\left(\frac{1}{M_{2}}-\frac{1}{M_{1}}\right)\dot{\bold{r}}$ is derived in \citet{2006epbm.book.....E} [see equation $(\ref{h2})$ in here and equation $(6.27)$ in page $251$ in \citet{2006epbm.book.....E}].\par
However, for the case of delta-function mass-transfer at periastron \citet{2006epbm.book.....E} derives the following secular time-evolution equation for the eccentricity 
\begin{equation}
\left\langle\frac{de}{dt}\right\rangle_{Egg}^{\delta}=
-2\dot{M}_{2}\left(\frac{1}{M_{2}}-\frac{1}{M_{1}}\right)(1+e)\label{eggletone}
\end{equation}
which has clearly a very different functional dependence on $e$ compared to our equation $(\ref{mye})$. Formula $(\ref{eggletone})$ has also been used in previous work, \citep[e.g.,][]{2000A&A...357..557S}. The reason for this discrepancy is that in these studies there is an error in the computation of the orbit-averaging integral $(\ref{avint})$ for a delta-function conservative mass-transfer in the case of stationary particles. In more detail \citet{2000A&A...357..557S} made a substitution $\cos{\nu}=1$ ($\nu =0$ at periastron) in equation $(\ref{egen})$ instead of performing the complete averaging integral $(\ref{avint})$ with delta-function mass-transfer at periastron.

\section{Non-conservative phase-dependent mass-transfer for point-masses (Bondi-Hoyle)}

Based on the Bondi-Hoyle accretion scenario \citep{1944MNRAS.104..273B} the accretion rate from a stellar wind can be written as
\begin{equation}
\dot{M}_{2}=a_{BH}\frac{4\pi\left(GM_{2}\right)^{2}}{\left(V_{rel}^{2} + c_{w}^{2}\right)^{3/2}}\rho_{w}
\end{equation}
where $a_{BH}$ is the Bondi-Hoyle accretion efficiency, $c_{w}$ is the wind sound speed and $\bold{V}_{rel}$ is the is the relative velocity of the wind with respect to the accreting star. An assumption commonly being made \citep[e.g.,][]{1998Icar..134..303D,2008AJ....135.1785J,2009ApJ...705L..81V,2012ApJ...759L..30P,2012ApJ...756..132S,2016RSOS....3.0571V} is that the density of the wind $\rho_{w}$ in the vicinity of the accretor follows the steady spherically symmetric profile $\dot{M}_{1}=-4\pi r^{2} \rho_{w} V_{w}$, where $\dot{M}_{1}$ is the wind-mass loss rate and $V_{w}$ is the wind velocity magnitude. Assuming that the wind flow is supersonic ($V_{rel} >> c_{w}$) the accretion rate $\dot{M}_{2}$ for a radial wind (with respect to the orbit) of speed $\bold{V}_{w}=V_{w}\bold{\hat{r}}$ is given by
\begin{equation}
\dot{M}_{2}=-a_{BH} \left(  \frac{GM_{2}}{r} \right)^{2}\frac{1}{V_{w}|V_{w}\bold{r}/{r}-\bold{\dot{r}}|^{3}}\dot{M}_{1}.\label{accmatter}
\end{equation}
where the efficiency factor for Bondi-Hoyle accretion $a_{BH}$ is in the range $0.5-1.0$ in the Bondi-Hoyle model \citep[e.g.,][]{1944MNRAS.104..273B,1985MNRAS.217..367S}, although it may be as low as $0.025$ in some specific cases \citep[e.g.,][]{1996MNRAS.280.1264T,2004NewAR..48..843E}.
\par For fast wind we have $V_{w}\gg V_{orb}$ and equation $(\ref{accmatter})$ reduces to
\begin{equation}
\dot{M}_{2}=|\dot{M}_{1}|\frac{r_{0}^{2}}{r^{2}}\label{fast}
\end{equation}
where all the constants have been summed up in $r_{0}$ to indicate dimensionality and $r_{0}$ is defined by
\begin{equation}
r_{0}^{2}=a_{BH}\frac{(GM_{2})^{2}}{V_{w}^{4}}
\end{equation}
\par For slow wind we have $V_{w} \ll V_{orb}$ and equation $(\ref{accmatter})$ reduces to
\begin{equation}
\dot{M}_{2}=|\dot{M}_{1}|\frac{r_{0}^{2}}{r^{2}}\left(\frac{V_{0}}{\dot{r}}\right)^{3}\label{slow}
\end{equation}
where again $V_{0}$ is a constant indicating dimensionality defined by $V_{0}=V_{w}$.
\par Both cases are non-conservative mass-transfer since only the fraction given by equation $(\ref{accmatter})$ is accreted by the companion. This fraction is phase dependent and in the case of fast wind is given by $\gamma(\nu)$, which is defined as $\dot{M}_{2}=-\gamma(\nu)\dot{M}_{1}$ and is given by 
\begin{equation}
\gamma(\nu)=\frac{r_{0}^{2}}{r^{2}}=\frac{r_{0}^{2}(1+e\cos{\nu})^{2}}{a^{2}(1-e^{2})^{2}}.\label{gamma}
\end{equation}
\par We then have $\dot{M}=(1-\gamma(\nu))\dot{M}_{1}$. Assuming point stationary masses (i.e., $\bold{b}=0$) and no extra angular momentum removal (i.e., $\zeta=0$), the total perturbing force for the phase-dependent partial fast wind accretion can be written as 
\begin{equation}
\bold{F}_{fw}=-\left(\frac{\dot{M}_{2}}{M_{WT}}+\frac{1}{2}\frac{\dot{M}}{M}\right)\bold{\dot{r}}
\end{equation}
where the subscript $fw$ stands for \emph{fast wind}  and for comparison we kept the same definition as in \citet{2006epbm.book.....E}
\begin{equation}
\frac{1}{M_{WT}}=\frac{1}{M_{2}}-\frac{1}{M_{1}}\label{fast2}
\end{equation}
Equations $(\ref{agen})$ and $(\ref{egen})$ then give for the \emph{secular semi-major axis and eccentricity evolution due fast wind Bondi-hoyle (BH) partial accretion in the adiabatic regime} the following equations
\begin{align}
\left\langle\frac{da}{dt}\right\rangle_{fw}^{BH}&=\frac{|\dot{M}_{1}|}{M}a-\dot{M}_{0}
 \frac{1+e^{2}}{1-e^{2}}a\left(\frac{2}{M_{WT}}+\frac{1}{M}\right)\label{fast3}\\
 \left\langle\frac{de}{dt}\right\rangle_{fw}^{BH}&=-\dot{M}_{0}
 e\left(\frac{2}{M_{WT}}+\frac{1}{M}\right)\label{fast1}
\end{align}
where we defined for comparison with \citet{2006epbm.book.....E} the quantity $\dot{M}_{0}$ as
\begin{equation}
\dot{M}_{0}=\frac{|\dot{M}_{1}|r_{0}^{2}}{a^{2}(1-e^{2})^{1/2}}.\label{correction}
\end{equation}
\par We mention here that in page $252$ and equation $(6.33)$ in \citet{2006epbm.book.....E}, there is an error and the factor $(1-e^{2})^{2}$ in the denominator should be $(1-e^{2})^{1/2}$ as in our equation $(\ref{correction})$. \par
After some algebra we can see that the sign of the right-hand-side of equation $(\ref{fast1})$ is determined by the sign of the quantity $2 q^{2}+q-2$ with a single zero solution in the range $q>0$ for $q=0.78$. For  $0<q<0.78$  the right-hand-side of equations $(\ref{fast3})$ and  $(\ref{fast1})$ is positive and thus for a mass-ratio $q$ in the range $0<q<0.78$, both the secular semi-major axis and eccentricity increase. For a mass-ratio $q>0.78$ the right-hand-side of equation $(\ref{fast1})$ is negative and the secular eccentricity always decreases while the sign of the right-hand-side of equation $(\ref{fast3})$ cannot be determined since it involves many and different parameters. This means that for a mass-ratio $q>0.78$ the semi-major secular evolution depends on many parameters including the orbital elements, the Bondi-Hoyle accretion efficiency parameter $a_{BH}$, the masses and the fast wind absolute velocity $V_{W}$.
\par The general case of slow wind is described by equation $(\ref{slow})$ and leads to a elliptical integral that demands numerical integration.\par
In the case that the wind velocity $V_{w}$ is of the same order like the orbital velocity $V_{w}\sim V_{orb}$, equation $(\ref{accmatter})$ must be employed. This case needs numerical manipulation. The same is true for the  any non-radial (with respect to the orbit) wind. However, assuming that the orbital velocity $\bold{V}_{orb}=\bold{\dot{r}}$ is always perpendicular to the radial wind which is always true in the circular case (Bondi-Hoyle circular approximation), then equation $(\ref{accmatter})$  simplifies to  
\begin{equation}
\dot{M}_{2}=-\frac{a_{BH}}{2} \left(  \frac{GM_{2}}{rV_{w}^{2}} \right)^{2}
\left[ \frac{1}{(1+V^{2})^{3/2}}\right]\dot{M}_{1}\label{circbondi}
\end{equation}
where we have assumed a constant velocity wind, a Bondi-Hoyle factor $a_{BH}$ and we defined
\begin{align}
V^{2}&=\frac{V_{c}^{2}}{V_{w}^{2}}\\
V_{c}^{2}&=\frac{G(M_{1}+M_{2})}{a}
\end{align}
where the subscript $c$ is for reminding the circular approximation made here.\par
\par Under these approximations equation $(\ref{circbondi})$ reduces to an equation similar to the case of fast wind equation $(\ref{fast})$ but with $r_{0}$ defined by
\begin{equation}
r_{0}^{2}=\frac{a_{BH}}{2} \left(  \frac{GM_{2}}{V_{w}^{2}} \right)^{2}
\left[ \frac{1}{(1+V^{2})^{3/2}}\right]\label{r0}
\end{equation}
Using $r_{0}$ defined in equation $(\ref{r0})$ equations $(\ref{fast3})-(\ref{correction})$ can be used accordingly to describe the Bondi-Hoyle accretion under the circular approximation. 
\par \citet{1988A&A...205..155B}, \citet{2000MNRAS.316..689K}, \citet{2002MNRAS.329..897H} and \citet{2015A&A...579A..49V} made use of equation $(\ref{circbondi})$ but they first substituted $1/r^{2}$ by its average $\left\langle 1/r^{2}\right\rangle=1/[a^{2}(1-e^{2})^{1/2}]$. With this approach, they forced the mass-transfer rate to be phase-independent, so performing the averaging integral $(\ref{avint})$ in equations $(\ref{agen})$ and $(\ref{egen})$ is oversimplified in their analyses. The right treatment involves orbit-averaging the full expression including the phase-dependent quantity $\gamma(\nu)$ given by equation $(\ref{gamma})$.
\par Moreover, in equation $(A.29)$ in \citet{2015A&A...579A..49V}  the last terms in the RHS of equations $(\ref{agen})$ and $(\ref{egen})$ are missing. These terms come from the change in total mass and orbital angular momentum of the system and cannot be neglected.

\section{Conclusions}

In this paper, we discussed the effect of various and different types of mass-loss and mass-transfer processes on the orbital evolution of eccentric binary systems. We made use of the general formulation and the phase-dependent time-evolution equations of orbital elements presented in Paper I. Based on them, we derived the secular time-evolution equations for the semi-major axis and the eccentricity under either the adiabatic approximation or the assumption of delta-function mass-loss/transfer at periastron.\par
We first discussed the cases of isotropic and anisotropic wind mass-loss. We then studied the non-isotropic ejection and accretion in a binary (including RLOF) in the conservative and non-conservative cases and for both point-masses and extended bodies. We presented specific examples of mass-loss/transfer under the point-mass approximation and compared them with similar work in the literature. We concluded discussing the case of phase-dependent mass accretion under the Bondi-Hoyle scenario.\\\\

\par The main results of the paper are summarized below:
\begin{itemize}
\item We studied the case of isotropic wind mass-loss. We proved that under the adiabatic approximation the secular osculating semi-major axis increases, while the secular osculating eccentricity remains constant. On the other hand, for delta-function isotropic mass-loss at periastron there is an increase in both the secular osculating semi-major axis and eccentricity.
\item We studied the case of anisotropic wind mass-loss. We showed that when the wind mass-loss is phase-independent the secular osculating semi-major axis is not affected. For wind mass-loss that is longitudinal independent and symmetrical about the equator, the secular osculating eccentricity remains constant. However, any asymmetry about the equator or any existing longitudinal dependence introduces changes in the secular eccentricity that depend on the specific spatial structure  of the wind on both latitude and longitude. Symmetric thin bipolar jets leave the secular eccentricity unchanged while any other asymmetrical or randomly directed jets induce secular eccentricity evolution.
\item We continued with the non-isotropic ejection and accretion in the case of point-masses. In this case:
\begin{enumerate}
\item A radial ejection from the donor with no accretion is secularly equivalent to the isotropic wind mass-loss case mentioned above.
\item  An ejection of mass in the opposite to the motion of the body direction leads to an increase in the secular semi-major axis and a decrease in the secular eccentricity of the orbit. An ejection in the same to the motion of the body direction leads to a semi-major axis secular decrease and an eccentricity secular increase. 
\item For rear ejection and delta-function mass-transfer at periastron,  both the secular semi-major axis and eccentricity increase. If the case of apastron, this will lead to an increase in the secular semi-major axis and a decrease to the secular eccentricity. For frontal ejection, delta-function mass-transfer at periastron will lead to a secular semi-major axis and eccentricity decrease, while delta-function mass-transfer at apastron will decrease the secular semi-major axis and increase the secular eccentricity.
\item In the case of isotropic/radial wind mass-loss from the donor and radial partial wind accretion and tangential mass-loss from the accretor, the secular semi-major axis decreases while the secular eccentricity increases.
\end{enumerate}
\item For the case of conservative or non-conservative RLOF, in contrast to tides which always act to circularize the orbit, mass-transfer may either increase or decrease the secular eccentricity, over timescales ranging from a few Myr to a Hubble time.
\item For stationary particles, the secular semi-major axis and eccentricity time-evolution depends on the mass ratio $q=M_{1}/M_{2}$. For adiabatic mass-loss rate, the secular eccentricity does not change while the secular semi-major axis decreases for $q>1$ and increases for $q<1$. In the case of delta-function mass-loss at periastron, both the secular semi-major axis and eccentricity decrease for $q>1$ and increase for $q<1$.
\item For phase-dependent mass-transfer assuming point and stationary particles, applying the general Bondi-hoyle accretion scenario requires numerical integration. However, the cases of fast wind or radial wind under the circular approximation are analytically integrable. For  $0<q<0.78$ both the secular semi-major axis and eccentricity increase. For a mass-ratio $q>0.78$ the secular eccentricity always decreases while the semi-major secular evolution depends on many parameters including the orbital elements, the Bondi-Hoyle accretion efficiency parameter $a_{BH}$, the masses and the fast wind absolute velocity $V_{W}$.
\end{itemize}
\par The secular time-evolution equations for the semi-major axis and the eccentricity we presented here for various and different cases of mass-loss/transfer could be used to model interacting eccentric binaries in star and binary evolution codes like StarTrack \citep{2008ApJS..174..223B}, BSE \citep{2002MNRAS.329..897H} or MESA \citep{2011ApJS..192....3P,2013ApJS..208....4P,2015ApJS..220...15P}.\par
In a future paper, we will attempt to relax the assumption of tidal locking in the case of extended bodies in the binary and study the effect of mass-loss and mass-transfer on the time-evolution of the binary orbital elements as well as the binary components spins. Assuming strong coupling between the spin and orbital angular momentum, we will derive secular time-evolution equations for the orbital elements and spins for the specific examples of mass-loss and mass-transfer studied in this paper including new applications.

\bigskip
\section*{ACKNOWLEDGEMENTS}
We want to thank our colleagues Dimitri Vears, Jeremy Sepinsky and Lorenzo Iorio who provided insight and expertise that greatly assisted the research. We are thankful to their comments that greatly improved the manuscript during its composition and revising stages.

\end{document}